\documentclass[%
 aip,
 amsmath,amssymb,
 reprint,%
]{revtex4-1}
\usepackage{graphicx}
\usepackage{dcolumn}
\usepackage{xcolor}
\usepackage{amssymb}
\usepackage{ulem}

\def \be{\begin{equation}}
\def \ee{\end{equation}}

\def \ba{\begin{eqnarray}}
\def \ea{\end{eqnarray}}

\def \bse{\begin{subequations}}
\def \ese{\end{subequations}}

\newcommand{\lw}[1]{\textcolor{black}{#1}}

\newcommand{\orcid}[1]{\href{https://orcid.org/#1}{\includegraphics[width=10pt]{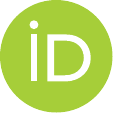}}}

\begin{document}

\title{Vane rheology of a fiber-reinforced granular material}

\author{Ladislas Wierzchalek}
\email{ladislas.wierzchalek@universite-paris-saclay.fr}
\affiliation{Universit\'e Paris-Saclay, CNRS, Laboratoire FAST, 91405 Orsay, France}

\author{Georges Gauthier}
\affiliation{Universit\'e Paris-Saclay, CNRS, Laboratoire FAST, 91405 Orsay, France}

\author{Baptiste Darbois Texier}
\affiliation{Universit\'e Paris-Saclay, CNRS, Laboratoire FAST, 91405 Orsay, France}

\date{\today}

\begin{abstract}
The addition of a small quantity of flexible fibers in a granular material is an efficient technique to increase the yield stress of the material. While the influence of fiber addition on the mechanical strength of granular media has been studied, much less is known about the flow properties of grain-fiber mixtures. In this article, we explore the effect of flexible fibers on the flow behavior of grain-fiber mixtures above the yield stress. We use a vane geometry to study the rheology of a dry granular material mixed with flexible fibers with different volume fractions and properties. The vane is immersed in the material, and the granular pressure increases with the depth of immersion. When the vane begins rotating, we observe a transient regime, which depends on the number of blades and is associated with the mobilization of material between the blades. Following this transient phase, a stationary regime is reached. By measuring and modeling the stationary flow that develops around the vane, we deduce the effective friction coefficient of the material from the torque measured on the vane. Following this approach, we investigate the effect of the fiber volume fraction and the aspect ratio on the effective friction coefficient of the grain-fiber mixture. Our results show that the effective friction coefficient increases linearly with fiber volume fraction and exponentially with fiber aspect ratio. These findings provide new fundamental insights into the flow properties of grain-fiber mixtures.

\end{abstract}

\maketitle

\section{Introduction}

Dry granular media are highly deformable materials and have relatively low mechanical strength, making them particularly vulnerable to erosion by wind, water currents, and other external forces. An inexpensive and efficient way to reinforce these materials is to add a small quantity of flexible fibers that entangle and stabilize the granular structure. This technique of fiber reinforcement is used in different contexts such as coast-line protection against sea erosion \cite{imran2022durability}, stabilization of sand dunes \cite{sharma2016behaviour}, equestrian sports surfaces \cite{hernlund2016sport} and natural construction materials such as adobes \cite{yetgin2008effects}. The vast areas of application of this technique have led to numerous studies within the field of soil mechanics \cite{shukla2017fundamentals}. These studies are based on measurements of the mechanical strength of fiber-reinforced soil samples in shear tests \cite{gray1983mechanics} and triaxial compression tests \cite{maher1990static,diambra2010fibre,ahmad2010performance}. A wide range of sands from various locations, along with fibers made from different materials and possessing diverse characteristics, have been tested using this approach. For example, Maher and Gray\cite{maher1990static} studied the response in compression of a Muskegon dune sand reinforced with glass fibers,  Diambra \textit{et al.}\cite{diambra2010fibre} investigated a Hostun RF sand reinforced with crimped polypropylene fibers and Ahmad \textit{et al.} \cite{ahmad2010performance} characterized the response of a silty sand reinforced with oil palm empty fruit bunch ﬁbers. In all these studies, the presence of fibers is observed to increase the strength of the material in compression. In their sand-fiber system, Diambra \textit{et al.}\cite{diambra2010fibre} observed that the net deviatoric strength of the sample increases by a factor two when 0.6\% of fiber content is added to the sand. On the contrary, they reported a limited effect of fiber presence on the resistance of the material to extension. To complement the experimental approach, numerical simulations using the discrete element method (DEM) have been developed to model the response of fiber-reinforced granular materials under compression and to explore the influence of parameters that are difficult to control in experiments, such as the effect of fiber orientation \cite{li2022fiber}. Fiber reinforcement techniques are also used in the preparation of concrete. Several studies have focused on the impact of fiber addition on the behavior of fresh concrete, a mixture of sand and concrete particles suspended in a liquid \cite{banfill2006rheology,hossain2012influence,abdelrazik2020effect,anas2022fiber}. Banfill \textit{et al.}\cite{banfill2006rheology} studied the influence of the addition of carbon fibers on the rheology of fresh concrete. They showed that the rheology fiber-reinforced concrete can be approximated with a Bingham model, where both the yield strength and the plastic viscosity increase with the fiber content and the aspect ratio.

Most of the work in the literature concerns the increase of the mechanical resistance of real soils constituted of highly polydispersed and irregular grains where fines, silt, and interstitial liquids are sometimes present that are mixed with natural or artificial fibers made of different materials and having sometimes a complex shapes (twisted or crimped). Moreover, most of these studies are limited to the slow deformation regime. All these grain-fiber systems do not constitute ideal models and in this study, we hypothesize that the investigation of a model granular material made up of spherical, monodisperse grains mixed with calibrated flexible fibers will provide additional insights into the role of the physical parameters that enhance the reinforcement of granular materials with fibers. We chose to study the response of this ideal mixture of grains and fibers in a rheological configuration with the aim of improving the prediction of the behavior of fiber-reinforced materials in various practical situations. 

The rheological characterization of granular materials without fibers has benefited from experiments conducted in various flow configurations, such as inclined planes and rotating drums \cite{jop2006constitutive,forterre2008flows}. These experiments have demonstrated that the rheology of granular materials is governed by the pressure exerted on the material. Consequently, specialized pressure-imposed rheometers have been developed to better characterize these materials \cite{boyer2011unifying,tapia2019influence}. Measurements made with these instruments have allowed to propose a semiempirical rheological law, known as the $\mu(I)$ rheology, which describes dense granular flows \cite{pouliquen2009non}. This rheology states that the effective friction of the material $\mu$ and its volume fraction $\phi$ both depend solely on the nondimensional shear rate, known as the inertial number $I$, which is expressed as $I =  \dot{\gamma}d / \sqrt{P/\rho}$ where $d$ is the grain diameter, $\rho$ their density, $\dot{\gamma}$ the flow shear rate and $P$ the pressure. For dry granular materials, the $\mu(I)$ function has been shown to increase from a starting value $\mu_1$ up to a limit value $\mu_2$ according to the empirical relation $\mu(I) = \mu_1 + (\mu_2 - \mu_1) / (1 + I_0 /I)$. The flow of granular materials has also been studied in less conventional configurations, such as the 'vane-in-cup' geometry\cite{ovarlez2011flows}, which has proven effective for characterizing yield stress fluids, including granular materials. Daniel \textit{et al.}\cite{daniel2008vane} studied the rheology of cohesionless glass beads in a vane geometry and analyzed the influence of the different geometrical parameters of the vane on the torque it experiences. Qi \textit{et al.}\cite{qi2020simulation} conducted DEM simulations in this configuration and demonstrated that the ﬂow features in the vane shear cell are equivalent to those in the classic annular Couette cell, allowing the determination of the granular material rheology.

In this paper, we adopt an approach similar to that which led to $\mu(I)$ rheology to investigate the response of model grain-fiber mixtures composed of spherical glass beads and flexible synthetic fibers using a vane-in-cup geometry. We examine the effects of fiber volume fraction and aspect ratio on the effective friction coefficient, which is derived from the torque applied to the vane. Section \ref{sec:setup} introduces the experimental setup, followed by section \ref{sec:Res}, which presents the rheological results of the grain-fiber mixture. Finally, we discuss the interpretation of these results in section \ref{sec:disc} before concluding.

\section{\label{sec:setup} Experimental setup}

\begin{figure}[h!]
\begin{center}
\includegraphics[width=0.9\linewidth]{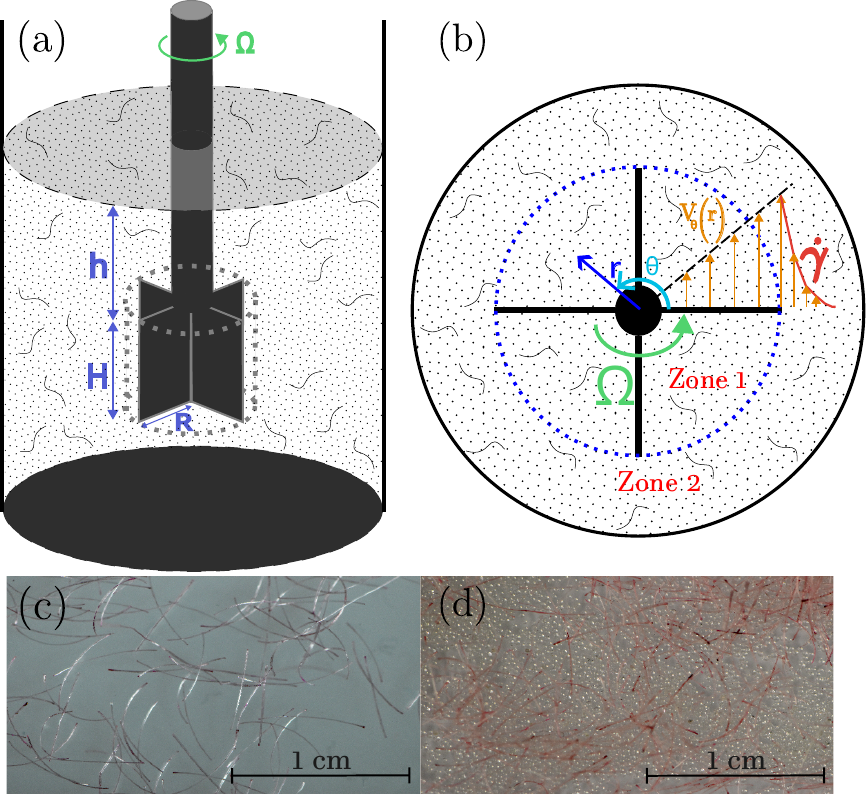}
\caption{$\rm{(a)}$ Sketch of the experimental setup used to explore the rheology of the fiber-reinforced granular material and the notations used in this study. $\rm{(b)}$ Top view of the vane geometry distinguishing two zones: zone 1 where the granular material is trapped between the blades, and zone 2 outside of the vane. $\rm{(c)}$ Fibers type B2, colored in red, alone. $\rm{(d)}$ Fibers type B2 mixed with glass beads.}
\label{fig:ExpSetUp}
\end{center}
\end{figure}

Experiments are performed with a vane tool rotating in a granular medium reinforced with fibers, see Fig. \ref{fig:ExpSetUp}(a). The vane tool is made from plastic (ABS) using 3D printing, a technique that enabled us to design vanes with different numbers of blades $N$, ranging from 3 to 6. The blades have a height $H=25.4$ mm and a radius $R=15.4$ mm and the shaft had a radius of $R_{\text{s}}=3.5$ mm.  The shaft of the vane tool is connected to an MCR 302 Rheometer (Anton Paar), which allows controlling the vane's vertical position and the rotation speed, $\Omega$, while measuring the torque, $\Gamma$, as it moves through the fiber-reinforced granular material. This instrument allows a constant rotation speed to be imposed within the range $\Omega = [0.06; 3.00]$ rad.s$^{-1}$ and measures torque between $50$ and $230$ mN.m. The advantage of the vane geometry for characterizing grain-fiber mixtures is that it allows a sample of this material trapped between the blades to move relative to the same material outside the blades, thus freeing the measurements from the influence of wall properties, which can have a significant effect in rheological measurements of granular materials \cite{hsiau2002stresses}.\\
The vane tool is immersed along the central axis of a cylindrical tank filled with glass beads mixed with flexible fibers. The cylindrical tank has a height $H_{\text{t}} =80$ mm and a radius $R_{\text{t}}=35$ mm which are much larger than the dimensions of the vane tool in order to avoid side and bottom wall effects on torque measurement \cite{seguin2008influence}. The vane tool is completely immersed in the grain-fiber material, with the top of the blades positioned at a depth $h$ below the free surface. In order to accurately set the depth of the vane in the medium, we use a protocol based on measuring the normal force experienced by the vane as it enters the medium, in order to detect the position of the free surface before lowering the vane to the desired depth. This procedure allows a precise control of the depth, $h$, with an accuracy of 0.3 mm. Note that the rotation of the vane is blocked during intrusion in order to limit disturbance to the initial state of the medium. The depth $h$ is also controlled at the end of each experiment to ensure that it remains the same at the beginning and end of the experiment. The granular medium is made of spherical glass beads of mean diameter $d=313 \pm 33~ \rm{\mu m}$ and of density $\rho=2475 \pm 25$ kg.m$^{-3}$ and will remain the same throughout this study. The volume fraction of the granular packing is $\phi=V_{\text{g}}/V_{\text{tot}}$ where $V_{\text{g}}$ is the volume of grains and $V_{\text{tot}}$ the total volume occupied by the packing. The flexible fibers added to the granular material are cylindrical filaments made of polypropylene (PP) or polyethylene terephthalate (PET) and have different lengths $l_{\text{f}}$ and different diameters $d_{\text{f}}$ that are summarized in table \ref{tab:fibersCarac}. Some of these fibers are shown in Fig. \ref{fig:ExpSetUp}(c) and in Fig. \ref{fig:ExpSetUp}(d). Before each experiment, fibers are added and mixed manually to the glass beads to reach a controlled volume fraction in fibers $\phi_{\text{f}}=V_{\text{f}}/V_{\text{tot}}$ where $V_{\text{f}}$ is the volume of fibers. In the literature, the preparation of fiber-reinforced soils in the presence of interstitial liquid has been done using the moist tamping method \cite{ibraim2007behaviour,diambra2010fibre}, the moist vibration method, a combination of both methods \cite{ibraim2012assessment,soriano20173d} or also an electric mixer \cite{banfill2006rheology,choobbasti2019shear}. For dry mixtures of fibers and grains, a manual mixing protocol has been employed \cite{ahmad2010performance,darvishi2018effect}, which has been shown to result in an approximately uniform distribution of fibers within the granular packing. In this study, we also utilized a manual mixing method and visually monitored the homogeneity of fibers distribution within the sample. We explore fiber volume fraction in the range of $[0;1.3]\%$ that remains below the maximal volume fraction of a random packing of fibers. The maximal volume of random packing of rigid fibers $\phi_{\rm{fc}}$ has been shown to be proportional to the inverse of the fiber aspect ratio \cite{philipse1996random} as: $\phi_{\rm{fc}}= 5.4 \, d_{\text{f}} / l_{\text{f}}$. For our fibers, this prediction gives a range of maximal fiber volume fraction, which is $ \phi_{\rm{fc}} \simeq [2.7;27]\%$. In practice, the maximal volume fraction of fibers that can be added to the grains is lower than the ideal limit established for pure fibers, and the homogeneity of the grain-fiber mixture is more difficult to achieve as this limit is approached. In this study, we investigate the rheology of grain-fiber mixtures in the limit of a low volume fraction of fibers where the homogeneity of the sample is good.\\
Since granular materials composed of particles of different sizes are prone to segregation \cite{gray2018particle}, we conduct preliminary tests using the vane geometry to assess fiber and grain segregation in this configuration. In practice, we dyed some of the fibers red and placed them in the granular medium near the vane at the center of the container ($r \lesssim R$ and $h \gtrsim z \gtrsim h+H $). The vane is then rotated for an extended period and we record when the red fibers first appear at the surface of the granular medium. No red fibers appear on the surface before 30 complete rotations of the vane. Furthermore, fiber segregation has no effect on the torque experienced by the vane, which remains constant over a large number of rotations (more than a few hundreds). In our experiments, we characterize the behavior of the grain-fiber material for a number of rotations smaller than 30 to avoid effects of segregation.\\
Additionally, the experimental setup includes a mirror positioned above the cylindrical tank at a angle $45^\circ$, along with a camera that records the movement of the glass beads on the surface of the tank when the vane tool is not fully immersed in the material. The film, captured at a frequency of 20 frames/s, is then analyzed using the Particle Image Velocimetry (PIV) method with the OpenPIV Python package to determine the velocity profile of the material at the surface of the tank.

\begin{table*}[htp!]
\vspace{1cm}
\begin{center}
     \begin{tabular}{|c|c|c|c|c|c|}
        \hline
        fiber's type & Material & Length $l_{\text f}$ (mm) & Diameter $d_{\text f}$ (µm) & aspect ratio $l_{\text f}/d_{\text f}$ & $\mathcal{D}$\\ 
        \hline
        A & PP & $6.0\pm0.5$ & $30\pm4$ & 200 & $1 \times 10^0$ \\ 
        \hline
        B1 & PP & $3.0\pm0.5$ & $100\pm11$ & 30 & $3 \times 10^{-3}$ \\
        \hline
        B2 & PP & $6.0\pm0.5$ & $100\pm11$ & 60 & $2 \times 10^{-2}$ \\
        \hline
        B3 & PP & $9\pm1$ & $100\pm11$ & 90 & $9 \times 10^{-2}$ \\
        \hline
        B4 & PP & $12\pm1$ & $100\pm11$ & 120 & $2 \times 10^{-1}$ \\
        \hline
        C & PET & $6\pm0.5$ & $330\pm27$ & 18 & $4 \times 10^{-4}$\\
        \hline
	\end{tabular}\\
	\caption{Materials and geometrical characteristics of fibers used in this study. The last column presents the values of the nondimensional parameter $\mathcal{D}$ characterizing fiber bending stiffness and defined in section \ref{sec:disc}. }
	\label{tab:fibersCarac}
\end{center}
\vspace{0.5cm}
\end{table*}

\section{\label{sec:Res} Experimental results}

Typical curves of the torque experienced by the vane when it starts to rotate in a granular material are shown in Fig. \ref{fig:Couple456pales}. We observe that the torque first undergoes a transient regime, reaching a maximum value before returning to a constant, stationary value. In the following, we will first analyze the transient regime and the influence of the number of blades. Subsequently, we will focus on the stationary regime and its dependence on the fiber volume fraction and the aspect ratio.

\subsection{Transient regime}

\begin{figure}[h!]
\begin{center}
\includegraphics[scale=0.55]{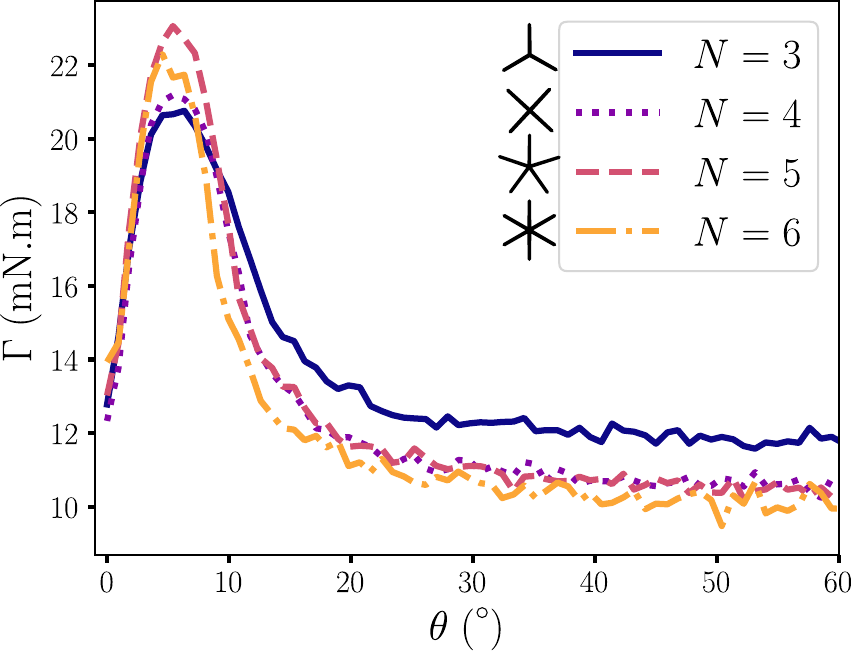}\\
\end{center}
\caption{Torque measured $\Gamma$ versus the rotation angle $\theta$ for vanes with different numbers of blades $N$. Experiments have been done with glass beads and fibers of type B2, $\phi_{\text{f}}=0\%$, $h=10$ mm and $\Omega=1.2$ rad.s$^{-1}$.}
\label{fig:Couple456pales}
\end{figure}

We return to Fig. \ref{fig:Couple456pales}, which shows the torque experienced by vanes immersed in grains ($\phi_{\text{f}}=0$) as a function of the rotation angle $\theta$. The different curves in this figure represent vanes with different numbers of blades ($N=3-6$) investigated for the same immersion depth $h=10$ mm and rotation speed $\Omega=1.2$ rad.s$^{-1}$. Each curves are the mean resulting from three different measurements obtained with the same experimental parameters. The typical standard deviation observed around the mean curve is of order 0.4 mN.m, which is less than 4\% in relative value. We observe that the torque starts from a nonzero value which depends on the history of the vane intrusion in the material, and then the torque increases up to a maximal value which is reached for $\theta \sim 10^\circ$ prior to decrease for larger $\theta$. For $\theta>50^\circ$ the torque has reached a constant and stationary value that depends little on the number of blades on the vane for $N>3$. Fluctuations in torque around the mean value may result from successive loading/rupture events in the granular medium, as observed in previous studies \cite{lehuen2020forces}. Conversely, the transient regime ($\theta<50^\circ$) is affected by the number of blades on the vane. We observe that as $N$ increases, the maximum value of the torque increases. Also, after reaching the maximum, the torque decreases more rapidly toward the stationary regime as $N$ increases. Note that we also conducted this experiment by replacing the vane with a full cylinder, and no torque overshoot was observed in the transient regime. However, we used the vane geometry as boundary conditions with the vane is more appropriate to grain-fiber mixtures.

\noindent In the following, we develop a model for the transient regime of the torque experienced by the vane. This model distinguishes two zones, the granular material inside the vane and that outside, zones 1 and $2$ respectively in Fig. \ref{fig:ExpSetUp}b. We begin by considering the contribution of zone 1. The transient regime of the drag force experienced by an object immersed in grains when it starts to move has been studied with numerical simulations \cite{seguin2019hysteresis}. It has been shown that the frictional contacts between grains require the intruder to travel a certain distance to be fully mobilized and for the drag force to reach its steady-state value. This transient regime has been successfully rationalized with a exponential relaxation term. We assume that each blade follows a similar transient regime before reaching the nominal force, and thus the total torque $\Gamma$ should evolve as $N~ \Gamma_{\text{b}} (1 - e^{-\theta/\theta_0})$ where $\Gamma_{\text{b}}$ is the nominal torque experienced by one blade and $\theta_0$ the characteristic angle to reach the stationary torque. The value of the parameter was measured for a rod oscillating in glass beads and was estimated\cite{darbois2021propulsion} to be $\theta_0 = 8^\circ$; thus, we assume in this model that $\theta_0$ is constant and equal to this value. However, in our case, an additional ingredient must be added to this scenario. When the granular material trapped between the blades enters a solid-body rotation with the blades, the drag force drops to zero due to the disappearance of relative motion between the granular material and the blades. We hypothesize that the solid rotation of the granular material between the blade is reached on a characteristic angle, which scales with the angle separating blades $2 \pi / N $. In order to account for this solid rotation limit, we add an exponential term to the previous relationship which becomes $N \Gamma_{\text{b}} (1 - e^{-\theta/\theta_0}) \, e^{-\alpha N \theta / 2 \pi}$ where $\alpha$ is a numerical coefficient. Then, we consider the contribution of the material sheared in the outer part of the vane (zone 2 in Fig. \ref{fig:ExpSetUp}b) in the transient regime. We assume that the transmission of the rotation motion imposed by the vane in the sheared layer also verifies exponential relaxation toward a stationary value, denoted as $\Gamma_\infty$, with a characteristic angle $\theta_0$. The torque resulting from the shearing of the outer layer (zone 2) is, therefore, taken into account by the expression $\Gamma_\infty (1 - e^{-\theta/\theta_0})$. Summing both contributions, we get the following expression for the total torque experienced by the vane as a function of the angle of rotation: 

\be\label{eq:model_transient}
\Gamma= N \, \Gamma_{\text{b}} (1 - e^{-\theta/\theta_0}) \, e^{-\alpha N \theta / 2 \pi} + \Gamma_\infty(1-e^{-\theta/\theta_0}) .
\ee

\begin{figure}[h!]
\begin{center}
\includegraphics[scale=0.55]{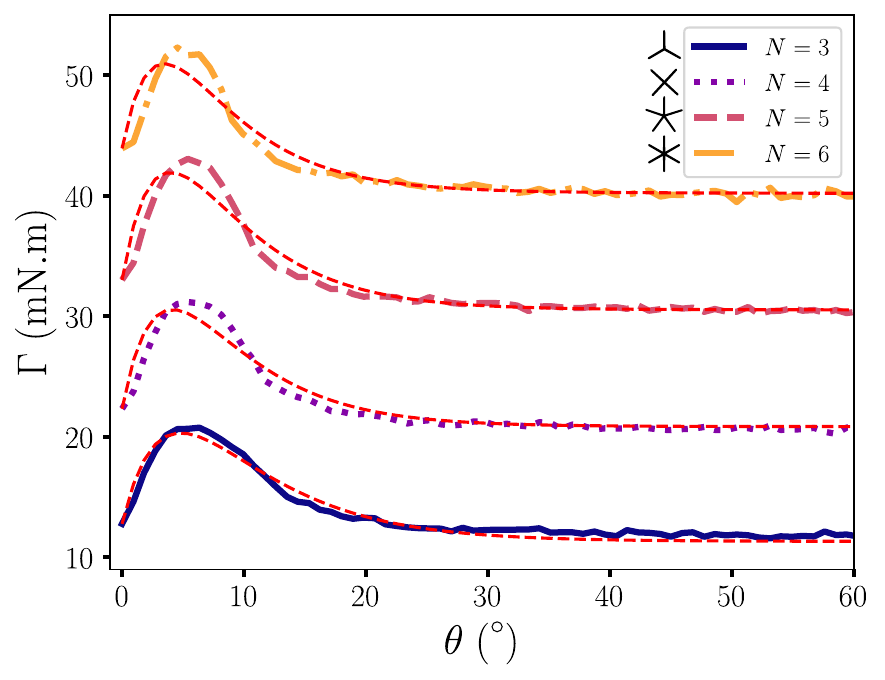}\\
\end{center}
\caption{Torque measured $\Gamma$ versus the rotation angle $\theta$ for vanes with different numbers of blades $N$ (solid lines) and best fits of the experiments (red dashed lines). For the sake of visibility, each curves contains an offset of $0$, $10$, $20$ and $30$ mN.m respectively. Experiments have been done with $\phi_{\text{f}}=0\%$, $\Omega=1.2$ rad.s$^{-1}$ and $h=10$ mm. The dashed lines show the best fits of the data with the model given by Eq. (\ref{eq:model_transient}) where $N \Gamma_{\text{b}}$ and $N \alpha$ have been considered as free parameters. In this fitting procedure, the transient angle $\theta_0$ is held constant at $\theta_0=8^\circ$.}
\label{fig:graph_ajust_transi_essai}
\end{figure}

\noindent In order to compare this prediction with our observations, we look for the best fit of the experimental data for $N=3,4,5,6$ with Eq. (\ref{eq:model_transient}) considering $N \Gamma_{\text{b}}$ and $N \alpha$ as free parameters, and $\Gamma_\infty$ is set to the mean value of the torque measured in the stationary regime (for $\theta>50^\circ$). The best fits of the data for different numbers of blades are presented in Fig. \ref{fig:graph_ajust_transi_essai} where the curves have been shifted vertically for the sake of visibility. We observe that the model reasonably captures the transient regime experienced by the vanes. 

\begin{figure}[h!]
\begin{center}
\includegraphics[scale=0.55]{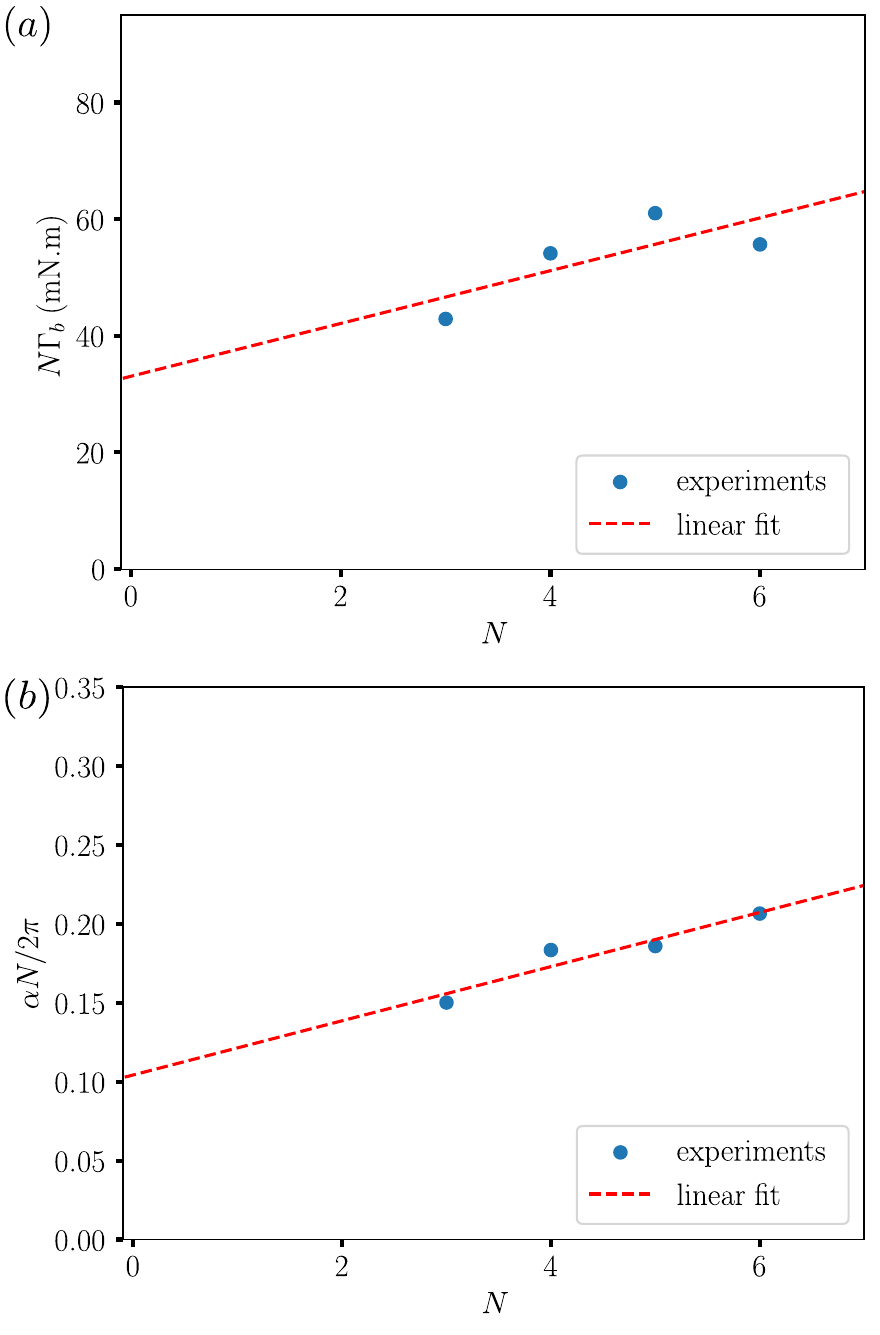}
\caption{Parameters $(a)$ $N \Gamma_{\text{b}}$ and $(b)$ $N \alpha$ deduced from the best fit of the data presented in Fig. \ref{fig:graph_ajust_transi_essai} as a function of the number of blades $N$. The red dashed lines indicate the linear fit of the data. These fits have been done for measurements made for $\phi_{\text{f}}=0\%$, $\Omega=1.2$ rad.s$^{-1}$ and $h=10$ mm.}
\label{fig:graph_ajust_transi_parametres}
\end{center}
\end{figure}

\noindent The parameters $N \Gamma_{\text{b}}$ and $N \alpha$ found for these fits are presented in Figs. \ref{fig:graph_ajust_transi_parametres} as a function of $N$. We observe that both parameters show a linear increase with $N$, which reinforces the validity of the model. The linear fit of $N \Gamma_{\text{b}}$ as a function of $N$ permits to estimate that $\Gamma_{\text{b}}= 4.5$ mN.m. This value can be compared with the estimation of the torque imposed by the granular pressure applying on a blade of surface $HR$ immersed at a depth $h$, \textit{i.e. } $\phi \rho g (h+H/2) H R^2 /2$ which gives $1$ mN.m in our case. This estimation and our measurements are of the same order of magnitude and agrees with the fact that prefactors in the expression of the granular forces are typically of order $5-10$ with a large dispersion \cite{faug2015macroscopic,seguin2018buckling}. Finally, the proposed model describes the transient regime of the torque experienced by the vane as it begins rotating within the granular material.\\

\subsection{Stationary regime}

\begin{figure}[h!]
\begin{center}
\includegraphics[scale=0.55]{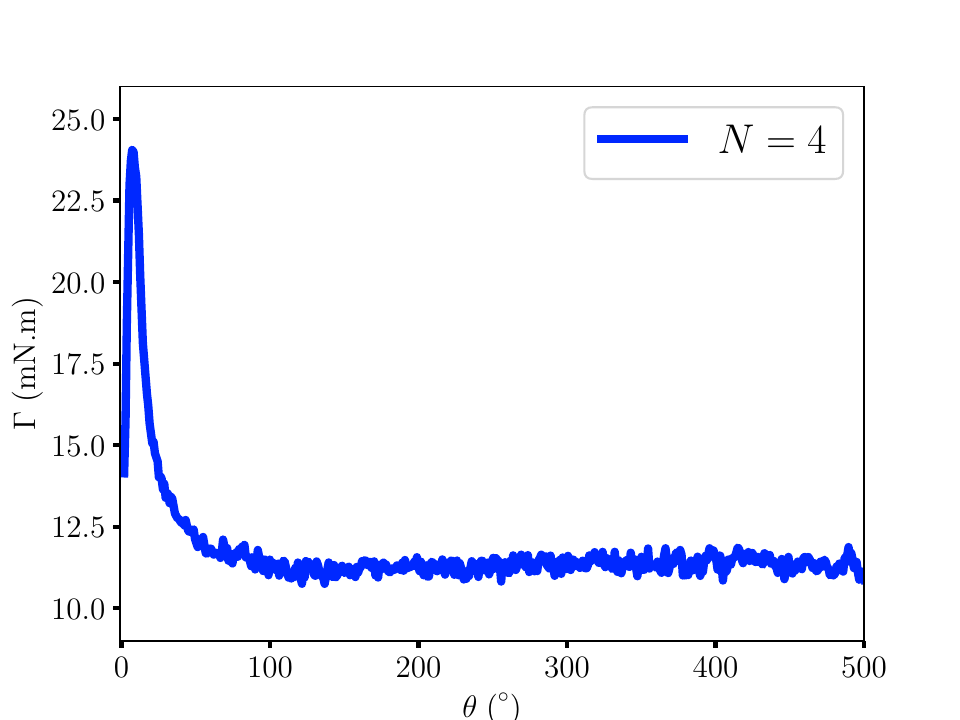}\\
\end{center}
\caption{Torque measured $\Gamma$ versus the rotation angle $\theta$ for vane with the number of blades $N=4$. Experiment has been done with fibers type B2, $\phi_{\text{f}}=0.15\%$, $h=10$ mm, and $\Omega=1.2$ rad.s$^{-1}$.}
\label{fig:graph_stationnary}
\end{figure}

In this section, we analyze the stationary value reached by the torque applied by the grain-fiber material on the vane after the transient regime. Figure \ref{fig:graph_stationnary} shows the evolution of the torque experienced by the vane immersed in a grain-fiber mixture for large $\theta$ values and highlights the existence of a stationary regime after the transient. The stationary value of the torque may depend on the geometry of the vane, its immersion depth, its speed of rotation and the properties of the grain-fiber material, which are to be deduced. To separate these dependencies, we first set the geometry of the vane and realize measurements with a four-bladed vane ($N=4$) throughout the remainder of the article. This choice is made to obtain measurements that are independent of the number of blades, while minimizing the normal force exerted on the vane during the first intrusion. We then consider the effect of the depth of the vane in the material. The influence of this parameter was studied by Daniel \textit{et al.} \cite{daniel2008vane} for a vane rotating in a dry granular material without fibers. In their work, they demonstrated that the torque experienced by the vane results from the shearing of the granular material on two surfaces: a disk at the base of the vane and a cylindrical surface at the outer edge of the vane. The total torque experienced by the vane is the sum of the lateral and basal torques, denoted as $\Gamma_{\rm side}$ and $\Gamma_{\rm bot}$, respectively. Each of these components is estimated from the hydrostaticlike pressure $P= \phi \rho g h$ confining the medium at a depth $h$, the effective coefficient of friction of the granular material $\mu$, and the corresponding surface area over which the pressure applies.  Following this approach, we can express the side and basal contributions as follows:

\be
\Gamma_{\rm side}= K \pi R^2 \mu \rho_{\rm g}g(H+h)^2
\label{eq:Gamma_side}
\ee

\be
\Gamma_{\rm bot}=\frac{2\pi}{3}R^3 \mu \rho_{\rm g}g(H + h)
\label{eq:Gamma_bot}
\ee

\noindent where $\rho_{\rm g}$ is the effective density of the medium ($\rho_{\rm g}=\phi \rho$) and $K$ is the Janssen coefficient, which accounts for the anisotropy of the stress distribution in a granular material. This coefficient has been determined experimentally and different values close to $1$ have been found \cite{prochnow2000dense,Erta__2001,bertho2003dynamical}. In order to verify if these predictions allow rationalizing the torque of a vane in a fiber-reinforced granular medium, we measure the stationary torque for different immersion depth $h$, keeping the rotational speed and the volume fraction of fibers constant. To verify the reproducibility of the experiments, we conducted them 15 times under identical conditions and demonstrated that the stationary torque could be measured with a relative accuracy of 2\%.

\begin{figure}[h!]
\begin{center}
\includegraphics[width=0.9\linewidth]{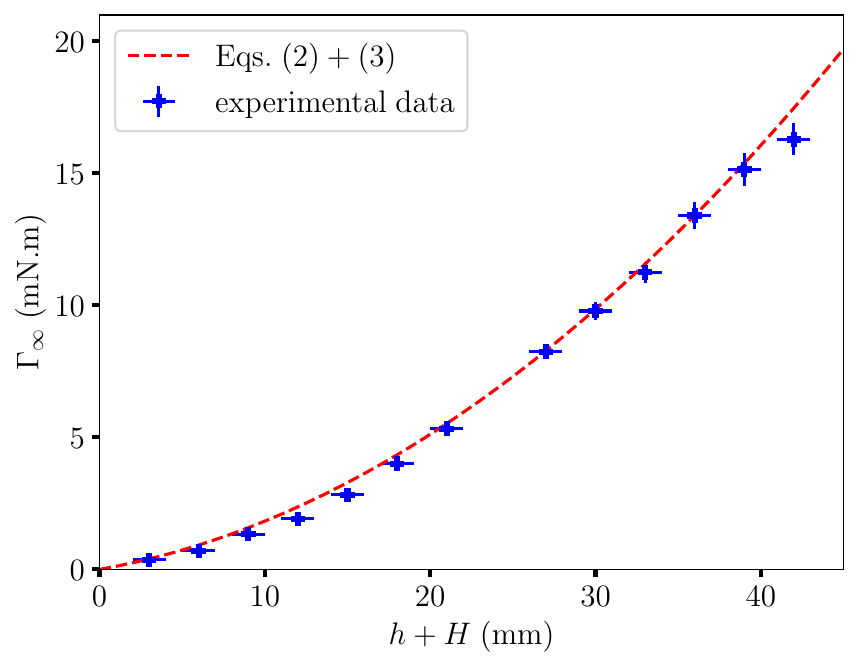}\\
\end{center}
\caption{Average stationary torque experienced by the four-bladed vane, $\Gamma_\infty$, as a function of the total immersed depth, $h+H$, for $\phi_{\text{f}}=0.15\%$ with fibers type B2. In each experiments, the vane rotates at a constant speed of $\Omega=1.2$ rad.s$^{-1}$. The red dashed lines show the best fit of the data with the sum of Eqs. (\ref{eq:Gamma_side}) and (\ref{eq:Gamma_bot}), considering $K$ and $\mu$ as free parameters. } 
\label{fig:Couple_profondeur_0}
\end{figure}

\noindent Figure \ref{fig:Couple_profondeur_0} shows the stationary torque $\Gamma_\infty$ as a function of the total immersion depth, $h+H$, for a grain-fiber mixture at $\phi_{\text{f}}=0.15$ \%. The supralinear increase of $\Gamma_\infty$ with the depth can be adjusted with the sum of Eqs. (\ref{eq:Gamma_side}) and (\ref{eq:Gamma_bot}) considering $\mu$ and $K$ as free parameters. The best fit of the data is represented with the red dashed line in Fig. \ref{fig:Couple_profondeur_0} and is obtained for $K=1.0 \pm 0.5$, which is in agreement with values determined previously in the literature where similar uncertainties have also been reported for this coefficient \cite{prochnow2000dense,Erta__2001,bertho2003dynamical}. In the following, we will assume that $K=1$. The value of the effective friction coefficient resulting from the fitting procedure is $\mu=0.74 \pm 0.20$ and is in the same range of the value determined for dry glass beads in a vane geometry by Daniel \textit{et al.} \cite{daniel2008vane}. Uncertainty calculated for both $K$ and $\mu$ comes from a Monte-Carlo method \cite{PAPADOPOULOS2001291}. We conclude that Eqs. (\ref{eq:Gamma_side}) and (\ref{eq:Gamma_bot}) can be used to deduce the effective friction coefficient of the material $\mu$ from the value of the stationary torque experienced by the vane. It should be noted that, in this approach, we have assumed that the granular material above the vane is also in solid rotation, similar to the material trapped between the blades. We have verified that changing this assumption regarding the flow above the vane results in only a slight change to the model predictions over the depth range investigated in this study. In addition, the direct visualization of the flow of material at the surface of the reservoir supports this hypothesis. 

\begin{figure}[h!]
\begin{center}\label{Profil_PIV}
\includegraphics[width=1\linewidth]{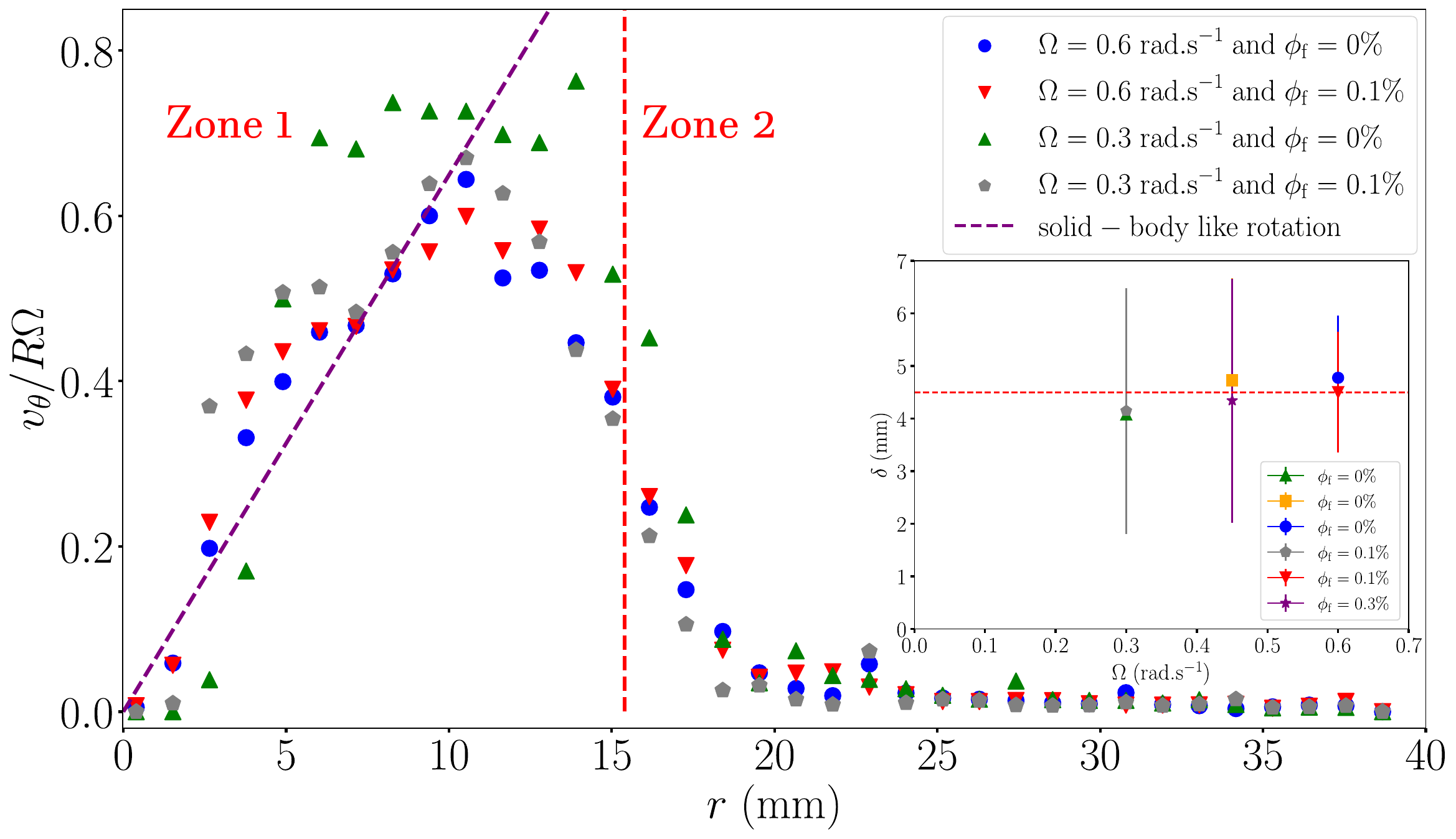}\\
\caption{Orthoradial velocity $v_\theta$ normalized by the blade tip velocity $R \Omega$ as a function of the distance $r$ to the central axis. Four experiments at different rotation speeds and fiber volume fractions are presented. Fibers used are the one referenced as type B3 in table \ref{tab:fibersCarac}. The purple dashed line represents the limit of solid-body rotation and the vertical red dashed line indicates the tip of the vane, $r=R$. \lw{In the inset, the shear layer thickness $\delta$, deduced from velocity profiles, is shown as a function of the vane rotation speed $\Omega$ for various fiber volume fractions.}}
\label{fig:Profil_PIV}
\end{center}
\end{figure}

\noindent Furthermore, we use the PIV method described in Section \ref{sec:setup} to investigate the surface flow of grains and fibers induced by the rotation of the vane in the quasistatic regime. This allows measurements of the velocity profile of the granular material that develops around the vane, see Fig. \ref{fig:ExpSetUp}b. These measurements would allow us to estimate the local shear rate of the flow $\dot{\gamma}= |{d v_\theta / dr}|$ and finally deduce the inertial number associated with this flow: $I= \dot{\gamma}d/\sqrt{P/\rho}$. Figure \ref{fig:Profil_PIV} presents the orthoradial velocity of the granular material $v_\theta$ normalized by the vane velocity $R \Omega$ as a function of the distance to the center of the vane $r$ for two different angular speeds and two different volume fractions of fibers. We observe that the normalized speed of the material first increases with $r$ up to a distance of about 12 mm, slightly less than the radius of the vane $R$, then saturates at a value of about 0.7 before decreasing to zero at larger distances. These observations correspond to the two zones highlighted in Fig. \ref{fig:ExpSetUp} : zone 1 where the material is in solid rotation with the vane and the orthoradial velocity increases linearly with the distance from the center and zone 2 where the material is sheared. Figure \ref{fig:Profil_PIV} also shows that the normalized velocity profile remains similar if the rotation speed of the vane or the volume fraction of fibers is varied. At the periphery of the vane, the granular material is sheared over a typical distance $\delta$ which is about 5 mm \lw{as it can be seen in the inset of Fig. \ref{fig:Profil_PIV}}, which corresponds to 16 grain diameters. In the following, we will assume that the typical distance $\delta$ over which the material is sheared at the periphery of the vane is constant and does not depend on the rotation speed, the volume fraction of fibers and the depth of the vane in the material. This assumption aligns with observations of granular flows in a cylindrical Couette configuration made in the limit of large gaps \cite{losert2000particle,reddy2011evidence,artoni2018shear}. These assumptions allow having a rough estimate of the shear rate experienced by the material as $\dot{\gamma} \simeq R \Omega / \delta$. Finally, we can estimate the inertial number $I$ of the flow around the vane, by considering that the pressure $P$ is the pressure resulting from the weight of the granular material at a depth corresponding to half the vane, $P= \rho_{\rm g} g (h + H/2)$.

\begin{figure}[h!]
\begin{center}
\includegraphics[width=1\linewidth]{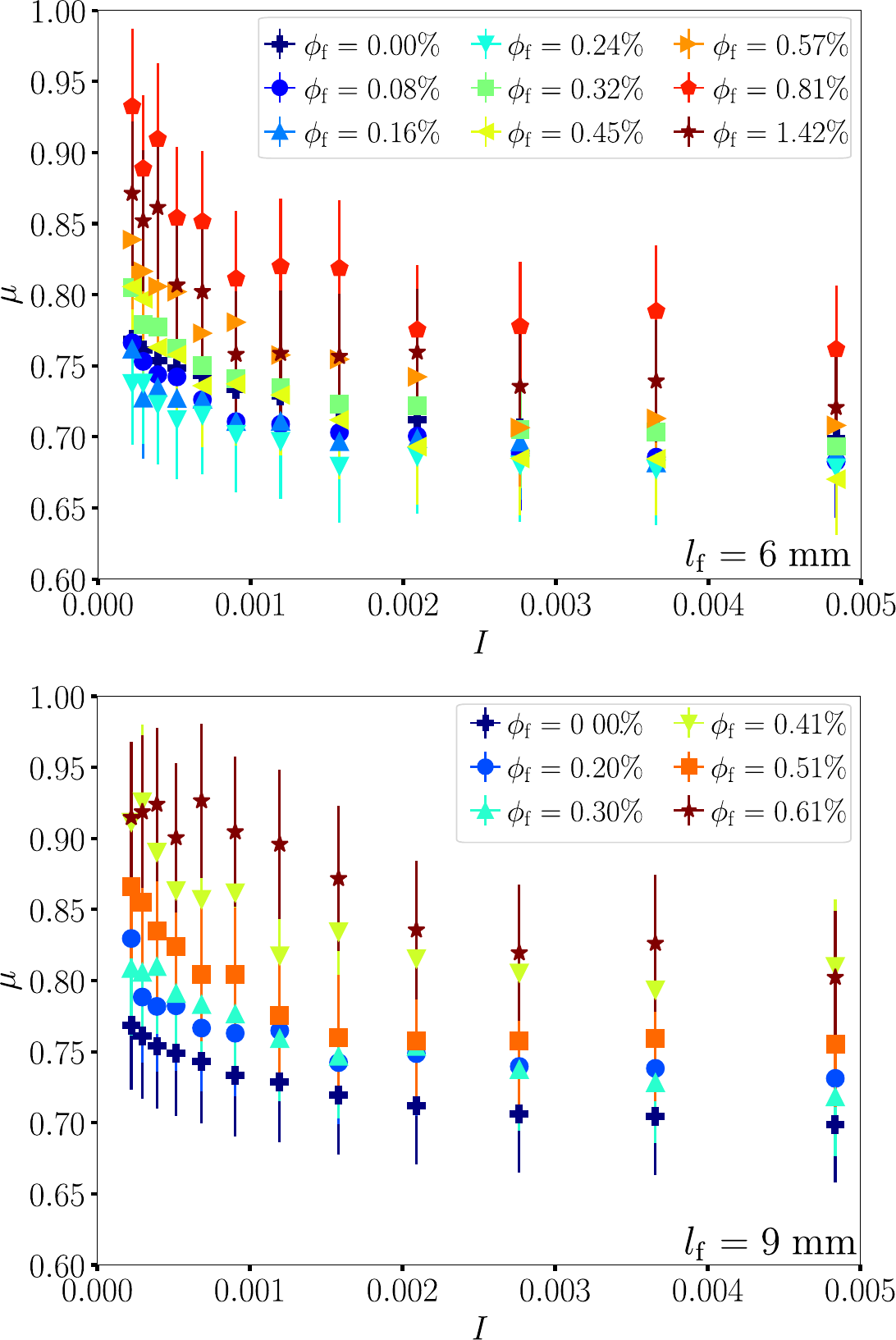}\\
\end{center}
\caption{Effective friction coefficient of a grain-fiber material $\mu$ as a function of the inertial number $I$ for different volume fraction of fibers $\phi_{\text{f}}$. Experiments are made with fiber types (a) B2 and (b) B3, with a four-blade vane immersed to a depth of $h=10$ mm.}
\label{fig:Graphe_µ(I)}
\end{figure}

\noindent With this approach, we deduce the inertial number of the flow $I$ from the speed of rotation of the blade and its depth of immersion, and the effective coefficient of friction $\mu$ of the material from the measurement of the torque experienced by the blade. We apply the method to measure the rheological properties of different samples of glass beads mixed with PP fibers, with $l_{\text{f}} = 6$ mm or $l_{\text{f}} = 9$ and $d_{\text{f}} = 100 ~\mu$m (types B2 and B3 in table \ref{tab:fibersCarac}) and a fiber volume fraction $\phi_{\text{f}}$ ranging from 0\% to 1.42\%. For each sample, the depth of the vane in the material is fixed at $h=10$ mm and we measure $\mu$ for increasing values of $I$, in the quasistatic regime. Figures \ref{fig:Graphe_µ(I)}(a) and \ref{fig:Graphe_µ(I)}(b) show $\mu$ as a function of $I$ for the range of $\phi_{\text{f}}$ studied in these experiments. In these plots, error bars correspond to the variance of that data estimated on 15 reproductions of the same experiment. We observe that the effective friction coefficient first decreases with $I$ before reaching a plateau value for $I>0.002$. Such a shear weakening behavior of the  $\mu(I)$-curve at small inertial numbers ($I<10^{-3} - 10^{-2}$) has already been reported for cohesionless grain rheology in both experiments and numerical simulations  \cite{kuwano2013crossover,degiuli2017friction,mandal2021rheology}. In the following, we will limit the scope of our study to the plateau value of $\mu$ for $I>0.002$. For these values of $I$, the medium is still in the quasistatic regime of flow. We observe in Figs. \ref{fig:Graphe_µ(I)}(a) and \ref{fig:Graphe_µ(I)}(b) that the effective friction coefficient $\mu$ increases with the volume fraction of fibers $\phi_{\text{f}}$, corresponding to a reinforcement of the material when the fiber volume fraction increases. Note that the increase in $\mu$ is significant even though the volume fraction of the fibers is low, less than 1.5 \%, and this increase can be seen in both fiber lengths in Figs \ref{fig:Graphe_µ(I)}(a) and \ref{fig:Graphe_µ(I)}(b). 

\begin{figure}[h!]
\begin{center}
\includegraphics[width=0.9\linewidth]{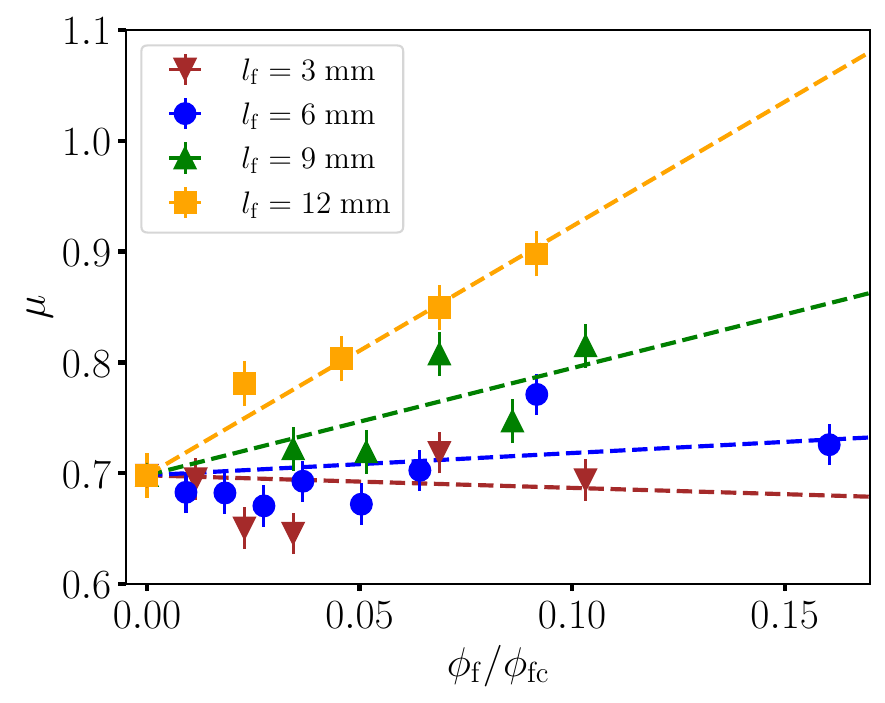}\\
\end{center}
\caption{Effective friction coefficient of a grain-fiber material $\mu$ as a function of the ratio of fiber volume fraction $\phi_{\text{f}}/\phi_{\rm fc}$ for four types of fibers (B1, B2, B3 and B4) with different lengths but the same diameter.}
\label{fig:Graphe_µ(phi)_longueur}
\end{figure}

\begin{figure}[h!]
\begin{center}
\includegraphics[width=0.9\linewidth]{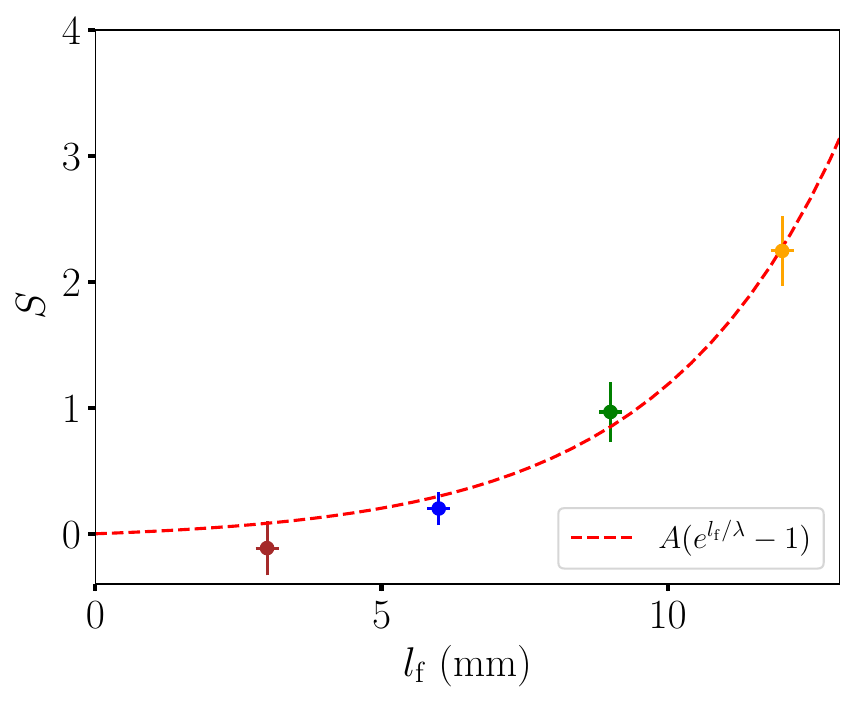}\\
\caption{Slope coefficient $S$ between $\mu$ and $\phi_{\text{f}}/\phi_{\rm fc}$ as a function of fiber length $l_{\text{f}}$ for the linear fit of the four sets of experiments shown in Fig. \ref{fig:Graphe_µ(phi)_longueur}. The solid line indicates the best fit of the data with the relation $S=A  ( e^{l_{\text{f}}/\lambda} - 1)$ where $A$ and $\lambda$ are considered free parameters and estimated to $A=0.05$ and $\lambda=3.2$ mm.}
\label{fig:Graphe_µ(phi)_longueur_modele_exp}
\end{center}
\end{figure}

\noindent In the following, we investigate how the effective friction coefficient in the plateau region depends on the fiber volume fraction, length, and diameter. For each volume fraction of fiber, we calculate the average value of the friction coefficient $\mu$ using data where $I > 0.002$. The error bars are determined from the one estimated on $\mu$ and assuming independent events. The mean friction coefficient $\mu$, derived from the measurements in Fig. \ref{fig:Graphe_µ(I)}, is shown as a function of fiber volume fraction in Fig. \ref{fig:Graphe_µ(phi)_longueur} (blue dots). To examine the influence of fibers length, we conducted similar experiments using grain-fiber mixtures with fibers of different lengths (types B1-B4 in Table \ref{tab:fibersCarac}), corresponding to lengths of 3, 6, 9, and 12 mm, all with the same diameter. The results are also plotted in Fig. \ref{fig:Graphe_µ(phi)_longueur} using different symbols. The data exhibit some dispersion, which could be attributed to uncertainties in the estimation of the immersion depth of the vane and small heterogeneities in the grain-fiber mixture. Since the maximum fiber volume fraction, $\phi_{\rm fc}$, depends on the fiber length, we plot the data in Fig. \ref{fig:Graphe_µ(phi)_longueur} as a function of the ratio $\phi_{\rm f}/\phi_{\rm fc}$. This allows comparison across different fiber lengths. For each data set, the effective friction coefficient increases with $\phi_{\text{f}}/\phi_{\rm fc}$, and the increase is more pronounced with longer fibers. Note that, for 12 mm fibers, the effective coefficient of friction increases by a factor of 1.5 for a fiber volume fraction as low as $\phi_{\text{f}}=1\%$. Each data set in Fig. \ref{fig:Graphe_µ(phi)_longueur} can be fitted with a linear relationship to derive a slope coefficient $S$. The variation of $S$ with fiber length, $l_{\text f}$, is presented in Fig. \ref{fig:Graphe_µ(phi)_longueur_modele_exp}. We observe a supralinear increase in the slope coefficient $S$ with $l_{\text{f}}$, highlighting the critical influence of fibers length on reinforcement. The trend can be modeled using the empirical relation $S = A(e^{l_{\text{f}}/\lambda} - 1)$, where $A$ and $\lambda$ are free parameters. The best fit of the data gives $A = 0.05$ and $\lambda = 3.2$ mm, which corresponds to approximately 10 grain diameters.

\noindent Based on these results, an empirical relationship can be proposed to account for the influence of flexible fibers on the effective friction coefficient of grain-fiber materials in the quasistatic regime. The proposed model can be expressed as

\begin{equation}
 \mu (\phi_{\text{f}},l_{\text{f}},d_{\text{f}}) = \mu(\phi_{\text{f}}=0) + A \left(e^{l_{\text{f}}/\lambda}- 1 \right) \frac{\phi_{\text{f}}}{\phi_{\text{fc}}}
 \label{eq:rheological_law}
\end{equation}

\noindent with $A=0.05$, $\lambda = 3.2~\rm{mm} \simeq 10\, d$ and reminding that $\phi_{\rm fc}=5.4~d_{\text{f}}/l_{\text{f}}$.\\

\begin{figure}[h!]
\begin{center}
\includegraphics[width=0.9\linewidth]{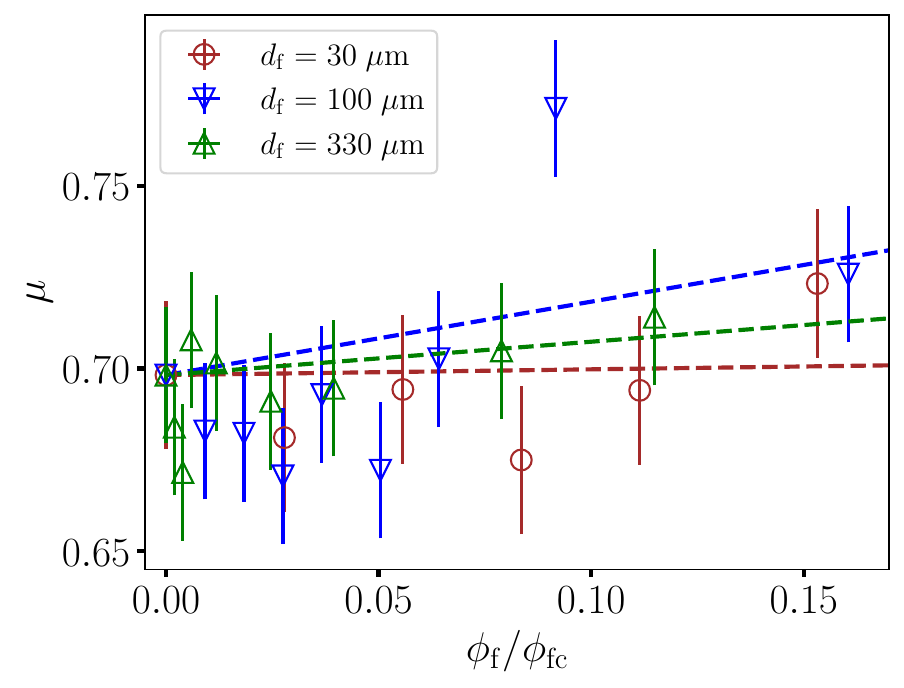}\\
\end{center}
\caption{Effective friction coefficient of a grain-fiber material $\mu$ as a function of the ratio of fiber volume fraction $\phi_{\text{f}}/\phi_{\rm fc}$ for three types of fibers with different diameters but the same length $l_{\rm f}=6 \, \rm{mm}$.}
\label{fig:Graphe_µ(phi)_diametre}
\end{figure}

\noindent To test the validity of Eq. (\ref{eq:rheological_law}), we explore the dependence of the effective friction coefficient $\mu$ with the fiber diameter $d_{\rm{f}}$. Following the same protocol and analysis as in the previous experiments, we use samples of grains mixed with fibers of the same length $l_{\text{f}}$, but with different diameters $d_{\text{f}}$ ranging from 30 to 330 $\mu$m (fiber types A, B2, and C in table \ref{tab:fibersCarac}). These results are plotted in Fig. \ref{fig:Graphe_µ(phi)_diametre} as a function of $\phi_{\text{f}}/\phi_{\rm {fc}}$. Despite the variability observed, we obtain a slight increase in the effective friction coefficient $\mu$ with increasing $\phi_{\text{f}}$ for all three fiber types studied. Notably, each data set follows a linear trend, and the slope of the trend is comparable across all fiber types, within the limits of experimental accuracy. This suggests that the effective friction coefficient $\mu$, when expressed as a linear function of the volume fraction ratio $\phi_{\text{f}} / \phi_{\rm {fc}}$, effectively captures the behavior across a wide range of fiber diameters varying by more than a factor of 10 in our experiments. Therefore, the dependence of Eq. (\ref{eq:rheological_law}) on fiber diameter $d_{\text{f}}$ does not appear to disagree with our observations.

\section{\label{sec:disc} Discussion}

In the approach adopted in this article, we interpret the increase in torque experienced by the vane due to the presence of fibers as an increase in the effective coefficient of friction. One might wonder whether the presence of fibers does not instead generate a geometric cohesion resulting from fiber interlocking with grains and each other. The concept of geometric cohesion was invoked by Dumont \textit{et al.}\cite{dumont2018emergent} to describe the behavior of a column made of beaded chains which no longer collapse under the effect of gravity when the chains are long enough. This concept is also used to understand the stability of a column made of star-shaped particles, which has been observed to increase with the number of branches of the star and the friction between the particles \cite{aponte2024geometric}. In the case of our grain-fiber systems, the presence of a geometric cohesion would lead to the appearance of a cohesive stress $\tau_{\text{c}}$ in the constitutive relation of the material linking the shear stress $\tau$ with the pressure: 
$\tau = \tau_{\text{c}} + \mu(I) P~$ 
\cite{mandal2020insights,deboeuf2023cohesion}. This cohesive stress would result in a shift in the relationship between shear stress and the pressure applied to the granular medium. In our study, this shift would be expected to manifest in the torque versus depth curves. In the presence of fibers, the experimental data in Fig. \ref{fig:Couple_profondeur_0} do not reveal any noticeable shift at zero pressure (\textit{i.e.}, for vanishing immersed depth). This suggests that, even if the presence of fibers in the grains leads to an effective cohesion of the medium, the cohesion remains weak within the investigated range of fiber volume fraction. This conclusion is consistent with the fact that, within the range of fiber volume fractions explored in this study, the presence of fibers does not allow for stabilizing a column of grains and fibers against gravity. These observations support our interpretation of the increase in torque as a change in the effective coefficient of friction in the fiber-grain mixture.

Granular materials are known to exhibit dilatancy when subjected to shear. The impact of fibers on this dilatancy behavior has been investigated in triaxial compression tests \cite{kong2019stress,szypcio2023stress}, revealing that the presence of fibers has a weak influence or even reduces the dilatancy of the grain-fiber mixture. For dry granular materials, the relative change in the volume fraction is linearly dependent on the inertial number \cite{tapia2019influence}, following the relation $\Delta\phi/\phi \simeq 0.1I$. In our study, which explores values of $I < 4 \times 10^{-3}$, this corresponds to a maximum relative variation of $\Delta\phi/\phi \simeq 4 \times 10^{-4}$. Consequently, dilatancy effects have a negligible influence on the torque exerted on the vane in our experiments and are too small to be detected by tracking the surface deformation of the material.

Previous findings on how the addition of fibers affects the macroscopic behavior of granular materials provide insights into the microscopic interactions between the grains and fibers. In particular, the exponential increase in the effective coefficient of friction with the fiber length (see Fig. \ref{fig:Graphe_µ(phi)_longueur_modele_exp}) is reminiscent of the sharp increase in the tensile force required to separate interleaved sheets of paper as the number of sheets increases \cite{alarcon2016self}. It also echoes the exponential increase in the tensile strength of an assembly of intertwined fibers with the angle of twist \cite{seguin2022twist}. Thus, our observations suggest that the flexible fibers become interleaved between the grains, creating a capstanlike effect \cite{johnson1987contact}, where the force required to pull the fibers increases exponentially with the length of contact between the fibers and grains, and consequently with the ratio $l_{\text{f}}/d$. In this scenario, the characteristic length $\lambda$ introduced in Eq. (\ref{eq:rheological_law}), is expected to increase linearly with the grain diameter, a prediction that should be confirmed by future studies. This scenario could also be validated through the development of numerical simulations that capture the dynamics of flexible fibers within a granular assembly, a task that has begun to be undertaken in the case of a small number of fibers in grains \cite{lobo2010fibre,bourrier2013discrete}. Last, the development of a model for the interaction between the fibers and the grains could take advantage of the models developed to rationalize the deformation of a flexible fiber penetrating into a 2D granular assembly \cite{algarra2018bending}, as well as in viscous \cite{marchetti2018deformation} and turbulent \cite{brouzet2014flexible} flows.

The last point to be discussed is the role of fiber flexibility on the flow properties of the grain-fiber mixture. The maximum deflection of the fibers can be estimated in the limit of linear elasticity. Thus, we consider a unique fiber as an elastic beam of Young modulus $E$, length $l_{\text{f}}$, with a cylindrical cross section of diameter $d_{\text{f}}$ whose quadratic moment of area is $\pi d^4_{\text{f}}/64$. The normalized deflection $w/l_{\text{f}}$ of this beam subjected to a uniform load $q$ can be written as \cite{audoly2000elasticity}
\begin{equation}
w/l_{\text{f}}= 64~q~l_{\text{f}}~d_{\text{f}} / (8 \pi E d^4_{\text{f}}).
\end{equation}
At a depth $h$ in the grains, the fiber is submitted to the granular pressure $\rho_{\rm g} g h$ which can deform it. Thus, the expected deflection of a flexible fiber is
\begin{equation}
w/l_{\text{f}}= ({4 \rho_{\rm g} g h}/{\pi E})~ (l_{\text{f}}/d_{\text{f}})^3.
\end{equation}
The deformation of a fiber in a granular medium is controlled by the nondimensional parameter $\mathcal{D}= ({\rho_{\rm g} g h}/{E})~ (l_{\text{f}}/d_{\text{f}})^3$. Small values of $\mathcal{D}$ ($\mathcal{D} \lesssim 10^{-3}$) correspond to the case of a rigid fiber, while large values of $\mathcal{D}$ ($\mathcal{D} \gtrsim 10^{-3}$) represent cases where the fiber is deformed and can entangle with the grains. The parameter $\mathcal{D}$ is estimated for each type of fibers and an immersion depth of $h=10$ mm, and these values are reported in the last column of table \ref{tab:fibersCarac}. For fibers used in this study, the parameter $\mathcal{D}$ varies in the interval $[4 \times 10^{-4},10^0]$. Fibers of type A and B are in the limit of flexible fibers (large $\mathcal{D}$) which can be bent by the pressure applied by the grains, and fibers of type C are in the limit of rigid fibers (small $\mathcal{D}$). The bending stiffness of fibers and their deformation in the granular material have a direct impact on the contact zone between the fibers and the grains. The role of the bending stiffness of an elastic beam in contact with a rigid cylinder has been considered by Grandgeorge \textit{et al.} \cite{grandgeorge2022elastic}. They studied the contact angle between the beam and the rigid cylinder as a function of the bending stiffness of the beam and the applied force. The contact angle increases monotonically as the normalized force, $\tilde{f} \propto \mathcal{D} / (l_{\text{f}}/d_{\text{f}})^2$, exceeds a threshold before saturating at $180^\circ$ for high load or low bending stiffness. The scenario mentioned in the previous paragraph, in which the fiber entangles the grains and induces a ‘capstan’ type effect, should be modulated by the effective contact surface between the fiber and the grains, which depends on the bending stiffness of the fibers. These physical ingredients will enable developing a microscopic model in the future to take account of the influence of fibers on the reinforcement of the granular medium.

\section{\label{sec:conc} Conclusion}

This study presents vane measurements of a dry granular material reinforced with flexible fiber and investigates the effects of fibers properties, specifically the volume fraction and the aspect ratio. These measurements allow us to deduce the effective friction coefficient of the grain-fiber mixture, showing that it increases linearly with the fiber volume fraction and exponentially with the fiber length. Based on these observations, we propose an empirical relation for the frictional rheology of grain-fiber materials in the limit of small inertial numbers. The effect of fibers on the macroscopic behavior of the reinforced granular material could be attributed to the interweaving of fibers between the grains and the increase in the fiber-grain contact surface with the fiber length. However, validating this scenario requires conducting numerical simulations to estimate the contact force between fibers and grains. Several aspects still need further investigation, including the distribution and orientation of fibers within the shear zone of the flow. Addressing this would require performing x-ray microtomography measurements of the grain-fiber material during vane rotation. Finally, it would be valuable to investigate the influence of confinement pressure on the behavior of grain-fiber mixtures, dilatancy effects, as well as to explore the effects at large inertial numbers and the impact of a viscous interstitial liquid.

\begin{acknowledgments}
The authors thank J. Amarni, A. Aubertin, L. Auffray, C. Manquest and R. Pidoux for their technical support. This work has been supported by the project FiLiGran ANR  22-CE30-0012. This work has benefited from fruitful discussions with H. Perrin and O. Pouliquen. 
\end{acknowledgments}

All the data are available in the repository http://doi.org/10.5281/zenodo.14905959.

%

\end{document}